\documentclass[prb,twocolumn,preprintnumbers,amsmath,amssymb]{revtex4}
\usepackage{graphicx}
\usepackage{color}

\newcommand{\muB}{\ensuremath{\mu_{\mathrm{B}}}}

\newcommand{\EF}{\ensuremath{E_{\mathrm{F}}}}
\newcommand{\dxy}{\ensuremath{d_{xy}}}
\newcommand{\dxz}{\ensuremath{d_{xz}}}
\newcommand{\dyz}{\ensuremath{d_{yz}}}
\newcommand{\dxxyy}{\ensuremath{d_{x^2-y^2}}}
\newcommand{\dzz}{\ensuremath{d_{z^2}}}

\begin{document}

\title{Exchange coupling in transition-metal nano-clusters on Cu(001) and Cu(111) surfaces}

\author{Phivos Mavropoulos}\email{Ph.Mavropoulos@fz-juelich.de} 
\author{Samir Lounis}
\author{Stefan Bl\"ugel}

\affiliation{Institut f\"ur Festk\"orperforschung (IFF) and Institute for Advanced Simulation (IAS), Forschungszentrum
  J\"ulich, D-52425 J\"ulich, Germany}

\begin{abstract}
  We present results of density-functional calculations on the
  magnetic properties of Cr, Mn, Fe and Co nano-clusters (1 to 9 atoms
  large) supported on Cu(001) and Cu(111). The inter-atomic exchange
  coupling is found to depend on competing mechanisms, namely
  ferromagnetic double exchange and antiferromagnetic kinetic
  exchange. Hybridization-induced broadening of the resonances is
  shown to be important for the coupling strength. The cluster shape
  is found to weaken the coupling via a mechanism that comprises the
  different orientation of the atomic $d$-orbitals and the strength
  of nearest-neighbour hopping. Especially in Fe clusters, a
  correlation of binding energy and exchange coupling is also
  revealed. 
\end{abstract}


\maketitle

\section{Introduction}

The magnetism of transition metal nanostructures on metallic surfaces
has been studied extensively in the past with emphasis on the magnetic
moments and ground state magnetic configuration. Sophisticated
experimental techniques for preparation, such as mass-selection and
soft-landing of free clusters, together with the ability to probe the
magnetism of these structures on the atomic scale, e.g. by X-ray
magnetic circular dichroism, have considerably advanced the
field.\cite{Lau02,Gambardella03} Strong fluctuations of the magnetic
properties have been found as a function of cluster size and shape,
position of the individual atoms in the cluster, or geometry of the
substrate.\cite{Lau02,Gambardella03,Wildberger95,Stepanyuk99,Izquierdo00,Spisak02,Lazarovits02,Mavropoulos06,Hafner07,Sipr07,Etz07,Robles08}
In some cases, though, it has been possible to recognize and interpret
regularities, and derive rules of thumb based on an understanding of
the electronic structure.

Supported clusters of atoms from the end of the $3d$-series (Fe, Co,
Ni) are known to be ferromagnetic, due to a double-exchange
mechanism. The spin moments in Fe clusters have been found to vary in
a regular way, namely the moment of each Fe atom scales down linearly
as a function of the number of Fe
neighbours.\cite{Mavropoulos06,Hafner07,Sipr07} This nice effect seems
to be independent of the substrate, and persists also in FeCo
clusters.\cite{Etz07} On the other hand, in pure Co clusters the
moment is practically saturated, while in Ni clusters the atomic
moments do not seem to correlate with the coordination number.

For V, Cr and Mn clusters, it has been found that the intra-cluster
coupling is antiferromagnetic, leading to frustration and
non-collinear magnetic order if the substrate provides appropriate
geometry (e.g., (111) surfaces).\cite{Bergman06,Bergman07} However, if
the substrate is magnetic, the exchange coupling to the substrate
atoms also plays a role and non-collinear magnetism can appear in Cr
and Mn clusters or chains on (001) Ni and Fe
surfaces.\cite{Lounis05,Lounis07,Lounis08a,Lounis08b}

The magnitude of the exchange coupling in $3d$ clusters on surfaces
has been less studied so-far. This is an important quantity, however,
in determining the stability of the ground state and the crossover or
blocking temperature. In addition, recent developments in experimental
techniques allow to probe the spectrum of magnetic excitations, e.g.,
via inelastic scanning tunneling
spectroscopy\cite{Hirjibehedin07,Otte08,Gao08,Balashov08} or
spin-polarized scanning tunneling spectroscopy.\cite{Meier08}

From the theory point of view, calculations on supported\cite{Minar06,Sipr07} and
free-standing\cite{Polesya07} transition-metal clusters have shown
that the inter-atomic exchange coupling fluctuates strongly with
respect of cluster size, shape, or position of the atoms in the
cluster. These works motivate an analysis of the local electronic
structure effects, in order to obtain a better insight into the
driving mechanisms for these fluctuations.

In this paper we perform such an analysis. We present
density-functional results on Cr, Mn, Fe, and Co nano-clusters (up to 9
atoms large) of mono-atomic height on Cu surfaces. For the
ferromagnetic clusters (Fe and Co) we also consider Cu(111) as
substrate, while for the antiferromagnetic ones we restrict our study
to collinear magnetism on Cu(001) (Cr and Mn show non-collinear
magnetism on Cu(111)). Our focus is on the inter-atomic exchange
coupling, where we attempt to find and understand the trends via the
details of the electronic structure. It is gratifying that, in some
simple geometries, the results can be interpreted in a transparent way
within a simple tight-binding model.

\section{Method of calculation}

Our calculations are based on density-functional theory within the
local density approximation (LDA), as parametrized by Vosko, Wilk, and
Nussair.\cite{Vosko80} The Kohn-Sham equations are solved in the
framework of the full-potential Korringa-Kohn-Rostoker Green-function
method (KKR) with exact treatment of the atomic cell shapes,
\cite{Papanikolaou02} using an angular momentum cutoff of $l_{\rm
  max}=3$. The Cu surfaces were modelled by slabs of a finite
thickness of 18 atomic layers; in all calculations, the LDA
equilibrium Cu lattice parameter (3.51~\AA) was used, while structural
relaxations were neglected. Within the KKR method, first the Green
function of the host (Cu surface) is calculated, while in a second
step a Dyson equation is solved for the Green function of the embedded
cluster. For all clusters, neighbouring host atoms were considered in
the self-consistent calculation in order to account for the screening
of the charge.

The exchange coupling is calculated by the method of infinitesimal
rotations.\cite{Liechtenstein87} This method is based on a hypothesis
of correspondence between the energy change $\Delta H$ of the Heisenberg
Hamiltonian
\begin{equation}
H=-\sum_{nn'}J_{nn'}\,\hat{e}_n\cdot\hat{e}_{n'}
\label{eq:2}
\end{equation}
and the total energy change found within DFT, $\Delta E_{\rm DFT}$,
upon rotating the magnetic moment directions $\hat{e}_{n}$ and
$\hat{e}_{n'}$. By virtue of the magnetic force theorem this energy
change can be calculated by the difference in the Kohn-Sham
eigenvalues. One obtains the exchange constants by taking a second
derivative,
$J_{nn'}=-\frac{1}{2}\,\partial^2H/(\partial\hat{e}_{n}\partial\hat{e}_{n'})
=-\frac{1}{2}\,\partial^2E_{\rm
  DFT}/(\partial\hat{e}_{n}\partial\hat{e}_{n'})$,
as \cite{Liechtenstein87}
\begin{eqnarray}
J_{nn'}&=& \frac{1}{4\pi}\int^{\EF} {\rm Im}\, {\rm Tr}_{lm}\,
G_{nn'}^{\uparrow}(E) \, [t^{\uparrow}_{n'}(E) -
  t^{\downarrow}_{n'}(E) ]
\nonumber \\
&&\times G_{n'n}^{\downarrow}(E)\, [t^{\uparrow}_n(E) -
t^{\downarrow}_n(E) ]\, dE
\label{eq:3} \\
&=& \int^{\EF}\! j_{nn'}(E)\, dE
\label{eq:4} 
\end{eqnarray}
where $G_{nn'}^{\uparrow,\downarrow}(E)$ is the inter-site KKR
structural Green function for spin up ($\uparrow$) or down
($\downarrow$), $t^{\uparrow,\downarrow}_{n}$ are spin-dependent
scattering matrices at sites $n$ and $n'$, and ${\rm Tr}_{lm}$
indicates a trace in angular momentum indices ($G_{nn'}$ and $t_{n}$
are matrices in angular-momentum space). We have also identified the
integrand as an ``exchange coupling density'' $j_{nn'}(E)$, which is
useful for the analysis of the results. In short, if the moments of
two atoms are rotated with respect to each other by a small angle
$\theta$, then the difference in energy arises primarily due to shifts
of the atomic energy levels or resonances. Then, $\theta^2j_{nn'}(E)$
gives the energy-shift of the local states at $E$, while
$\theta^2J_{nn'}$ gives the change in total energy; a subtraction of
the interaction with the rest of the magnetic atoms is also included
implicitly.  We also introduce the sum of the exchange interactions in
the $n$-th atom in the cluster,
\begin{equation}
J_0^{(n)} = \sum_{n'} J_{nn'},
\label{eq:5}
\end{equation}
which is a measure of ``local spin stiffness''. Integrating $j$ up
to a certain energy $E$, instead of $\EF$, yields $J(E)$ and $J_0(E)$
that also helps to analyze the contribution of specific states.

In practice, the set of equations (\ref{eq:2}) and (\ref{eq:3}) is an
accurate parametrization of the DFT total energy only for small
deviations from a reference state. Usually the ground state is chosen
as reference, which can be ferromagnetic or
antiferromagnetic. Eq.~(\ref{eq:3}) then yields $J>0$ if energy must
be payed for mutually rotating the spins direction, and $J<0$ if
energy is gained. Thus, starting from a ferromagnetic or
antiferromagnetic state within the LDA, $J>0$ always indicates
stability of this state, irrespectively if it is ferro- or
antiferromagnetic; the extra $(-)$ sign associated with
antiferromagnetism in (\ref{eq:2}) is absorbed, so to say, in either
$\hat{e}_{n}$ or $\hat{e}_{n'}$.

\section{Atomic spin moments}

\begin{figure}
\begin{center}
\includegraphics[width=8cm]{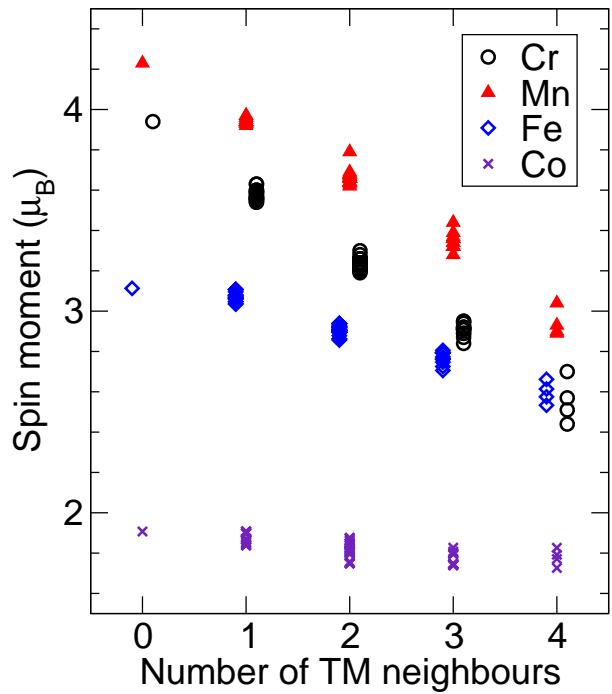}
\caption{Atomic spin moment of Cr, Mn, Fe, and Co in corresponding
  clusters of 14 different cluster geometry types, ranging from 1 to 9
  atoms, on the Cu(001) surface. The moment shows a linear trend as a
  function of the coordination number of nearest transition-metal (TM)
  neighbours. \label{fig:moments}}
\end{center}
\end{figure}

In Fig.~\ref{fig:moments} we show the dependence of the atomic moments
as a function of neighbouring atoms for Cr, Mn, and Fe, and Co.  The
strongest atomic spin moments are found, as expected, in Mn clusters,
ranging from 4.23 \muB\ for a single adatom to 2.89 \muB\ for the
central atom of a 9-atom cluster. Cr shows somewhat lower moments,
then comes Fe and finally Co. In a previous paper \cite{Mavropoulos06}
we showed that the atomic moments $M$ in Fe drop with increasing
coordination following approximately a simple linear relation,
\begin{equation}
M(i)=-aN_c(i) + b, 
\label{eq:1}
\end{equation}
where $N_c$ is the number of nearest neighbours of the $i$-th Fe atom
in the cluster and $a$ and $b$ are positive constants (independent of
$i$). This relation simply expresses the common wisdom that the moment
decreases with hybridization, and that the hybridization mainly
depends on the number of nearest neighbours. However, this was found
not to be so simple for Co, where the moment is more saturated and for
Ni.\cite{Mavropoulos06} Additional effects come from interference
among the $d$ states of remote atoms in the cluster, at the onset of a
narrow $d$-band formation at the Fermi level \EF, giving a stronger
relative scatter to the data.

\begin{figure}
\begin{center}
\includegraphics[width=8cm]{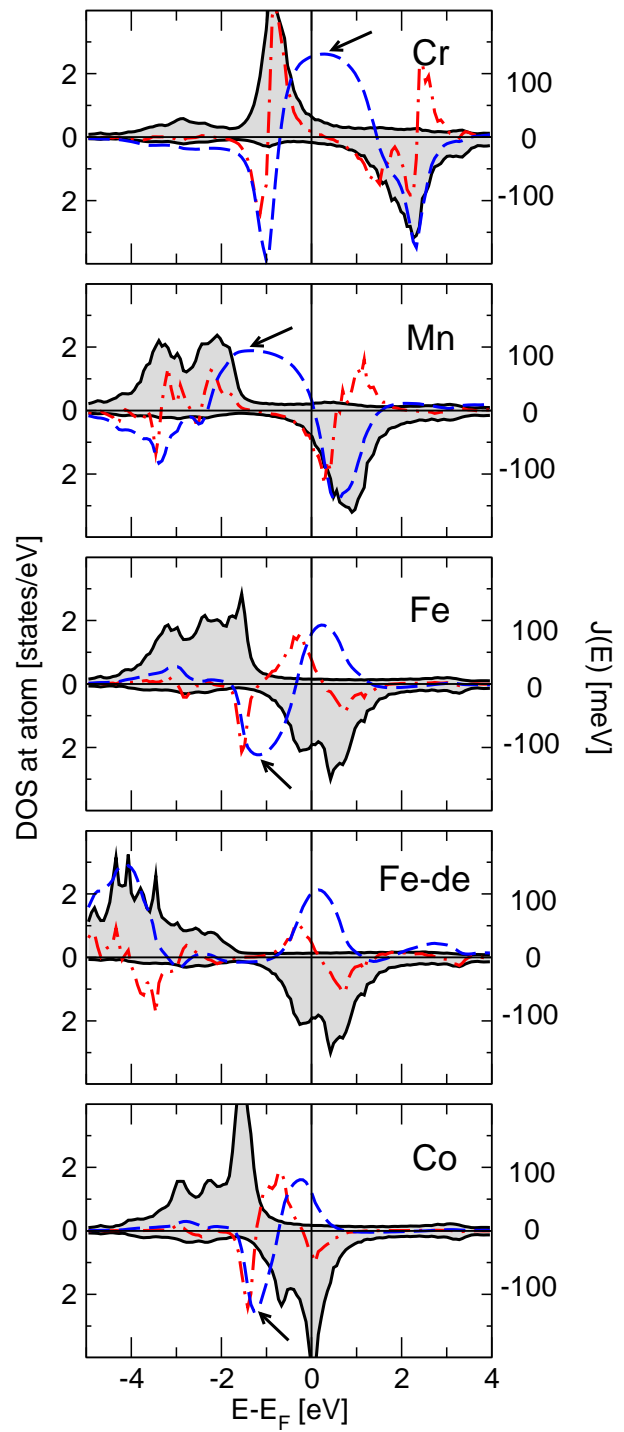}
\caption{Density of states and exchange coupling for Cr, Mn, Fe, and
  Co dimers on Cu (001). Label ``Fe-de'' corresponds to a Fe dimer
  with the majority states pushed to low energies, so that the double
  exchange becomes evident. Full, black line over shaded area (left
  scale): density of states; upper panels correspond to majority spin,
  lower panels to minority spin. Dashed, blue line (right scale):
  exchange coupling $J(E)=\int^{E}j(E')dE'$ as a function of energy;
  $J(\EF)$ corresponds to the actual exchange constant. Arrows
  indicate the particular local maxima of $J(E)$ that correspond to
  the kinetic-exchange mechanism (absent in the ``Fe-de'' case, where
  both maxima of $J(E)$ correspond to the double-exchange
  mechanism). Dashed-dotted red line: exchange density $j(E)$ (not to
  scale). \label{fig:dos}}
\end{center}
\end{figure}

In the case of Cr and Mn clusters, relation (\ref{eq:1}) holds rather
well (here for the absolute value of $M$, since the clusters are
antiferromagnetic). As the $d$-states of these elements are either
below the Fermi level (for majority-spin) or above the Fermi level
(for minority-spin), there is no $d$-band formation at \EF\ (see
Fig.~\ref{fig:dos} for the density of states of Cr, Mn, Fe, and Co
dimers). In addition, the antiferromagnetic configuration allows for
$d$-$d$ hybridization only between occupied and unoccupied states
making it comparatively weak, which is reflected on the smooth,
Lorenzian-like, minority-spin density of states, contrary to Fe and
Co. (The majority-spin resonances appear wider in Mn, Fe, and Co,
because of their interaction with the Cu substrate $d$-band that lies
at approximately the same energy.) Since the hybridization at \EF\
arises primarily due to the extended $s$ states of the first
neighbours, the behavior of the moments correlates smoothly with the
coordination. In the case of Co, the moment is almost saturated, changing
very little with coordination; the fluctuation in $M$ cannot be
correlated to $N_c$. The weak scatter of data for any particular
coordination corresponds to different atoms on different clusters.

\section{Exchange coupling: Preliminary remarks \label{sec:tb}}

We proceed with the exchange coupling in the nano-clusters. Before
presenting the density-functional results, we discuss how one can
interpret these in terms of a simple model.

The inter-atomic exchange coupling depends on various factors. The
nearest-neighbour exchange can be interpreted in terms of the local
density of states (DOS). The longer-range interactions, on the other
hand, are governed by a generalization of the
Ruderman-Kittel-Kasuya-Yosida (RKKY) interaction,\cite{Pajda02} depend
on the properties of the Fermi surface, and are far more complex in
nano-clusters, as a Fermi surface is not yet formed. Also, details of
the surroundings of the cluster can play a role due to quantum
confinement.\cite{Brovko09} Furthermore, the substrate below the
surface layer, e.g. buried magnetic clusters or layers,\cite{Brovko08}
can affect the long-range interactions; even non-magnetic buried
structures can play a role as spin-dependent wavefunctions incident
from a magnetic atom at the surface can be reflected at the buried
structure and propagate to another atom back on the surface,
especially in the presence of Fermi-surface focusing.\cite{Weismann09}
Another type of exchange interaction is the anisotropic exchange of
the Dzyaloshinskii-Moriya type, which has been found, for example, to
produce a long-wavelength, non-collinear magnetic ground state in Mn
overlayers on W(110).\cite{Bode07} This, however, depends on the
spin-orbit coupling, and even in transition-metal clusters on Au and
Pt (where the spin-orbit coupling is strong), the anisotropic exchange
is found to be two orders of magnitude weaker than the
nearest-neighbour Heisenberg exchange.\cite{Antal08,Mankovsky09} In
Cu, spin orbit coupling is much weaker, therefore we consider that its
effect can be safely neglected for the small-size clusters that we examine
here. Following these considerations, we focus our discussion only on
the nearest-neighbour Heisenberg-type exchange coupling, which is also
much stronger than more-distant interactions.

As is long known,\cite{Alexander64} there are two main contributions
to $J$, of different origin: the first, sometimes called ``kinetic
exchange'', favours antiferromagnetism in Cr and Mn. The second,
sometimes called ``double exchange'', favours ferromagnetism in Co and
Fe. Both mechanisms are associated with direct hopping among $d$
states of neighbouring atoms. As we will see below, the two mechanisms
can be present simultaneously and compete, reducing the absolute value
of $J$.\cite{Akai98}

To proceed with the discussion, consider two neighbouring transition
metal atoms. \emph{Kinetic exchange} arises from a level repulsion
between occupied majority states of an atom with unoccupied minority
states of a neighbouring atom, when the moments of the two are
oppositely oriented. In a simple tight-binding picture, let us assume
that the energy of the occupied $d$ level is at $E_d-\Delta<0$, while
the unoccupied level is at $E_d+\Delta>0$, where $2\Delta$ is the
exchange splitting; the Fermi energy is assumed to be at
$\EF=0$. Allowing for a hopping $t$ between two neighbouring atoms, the
levels move to $E_{\pm} = E_d \pm \sqrt{\Delta^2+t^2}$. Upon forcing
the two moments to a ferromagnetic configuration, the shift vanishes
and the occupied levels move higher, which costs energy $2[(E_d
-\Delta) - E_{-}] =2(\sqrt{\Delta^2+t^2}-\Delta)\approx t^2/\Delta$,
for $\Delta \gg t$ (the factor 2 accounts for the number of
occupied states). This picture does not change even in the presence of
a broadening by a hybridization $\Gamma$ of the levels with a
background continuum (as long as the broadened resonance does not
cross \EF), since the center-of-mass of the $d$ levels is normally not
shifted by $\Gamma$.

\emph{Double exchange} arises if the majority-spin or minority-spin
states of the two transition metal atoms are in the proximity of \EF,
i.e., either $E_d - \Delta\approx 0$ or $E_d+\Delta\approx 0$. Suppose
that the two moments are ferromagnetically oriented, and that
$E_d+\Delta= 0$. Allowing for a hopping $t$ between the $d$-states,
the minority levels split into bonding and antibonding, and are driven
to $E_{\pm} =\pm t$; $E_-$ is occupied, $E_+$ empty. Upon forcing the
two moments to an antiferromagnetic configuration, the splitting
vanishes, and the occupied $E_-$ move back to \EF. This costs energy
$t$. (The same splitting occurs also for the majority levels, but does
not affect the total energy as they are fully occupied and their
center-of-mass does not change). The energy gained by the
double-exchange mechanism, however, is reduced if the $d$-states are
resonant with a background hybridization $\Gamma$, because the tails
of the bonding and antibonding resonances cross \EF, so that their
repopulation partly counter-acts the energy gain. A simple calculation
shows that the energy gain by the hopping, $\int_{-\infty}^0 E\,\Delta
n(E)\,dE$, is a decreasing function of $\Gamma$ ($\Delta n(E)$ is here
the difference between the density of states without and with hopping,
and only the single-particle energies are considered).

Now, if the double-exchange mechanism is present, it must coexist with
kinetic exchange, which cannot be switched off. There is then a
competition between the two, and in the simple picture given above
(disregarding the hybridization), the criterion for a ferromagnetic
ground state is $t>t^2/\Delta$, i.e. $\Delta>t$. Otherwise the ground
state is antiferromagnetic. This analysis also suggests a way to
reveal the strength of the double-exchange mechanism in a calculation,
which we use below. If the majority-spin states are artificially
driven to lower energies by acting on them with an attractive
potential, the kinetic exchange $t^2/\Delta$ will decrease and what is
left will be mainly the double exchange part.

To conclude this subsection, nearest-neighbour exchange coupling can
partly be understood in terms of bond formation, but there are
competing factors that decide its final value.

\section{Exchange coupling: Results and discussion}

\begin{figure*}
\begin{center}
\includegraphics[width=16cm]{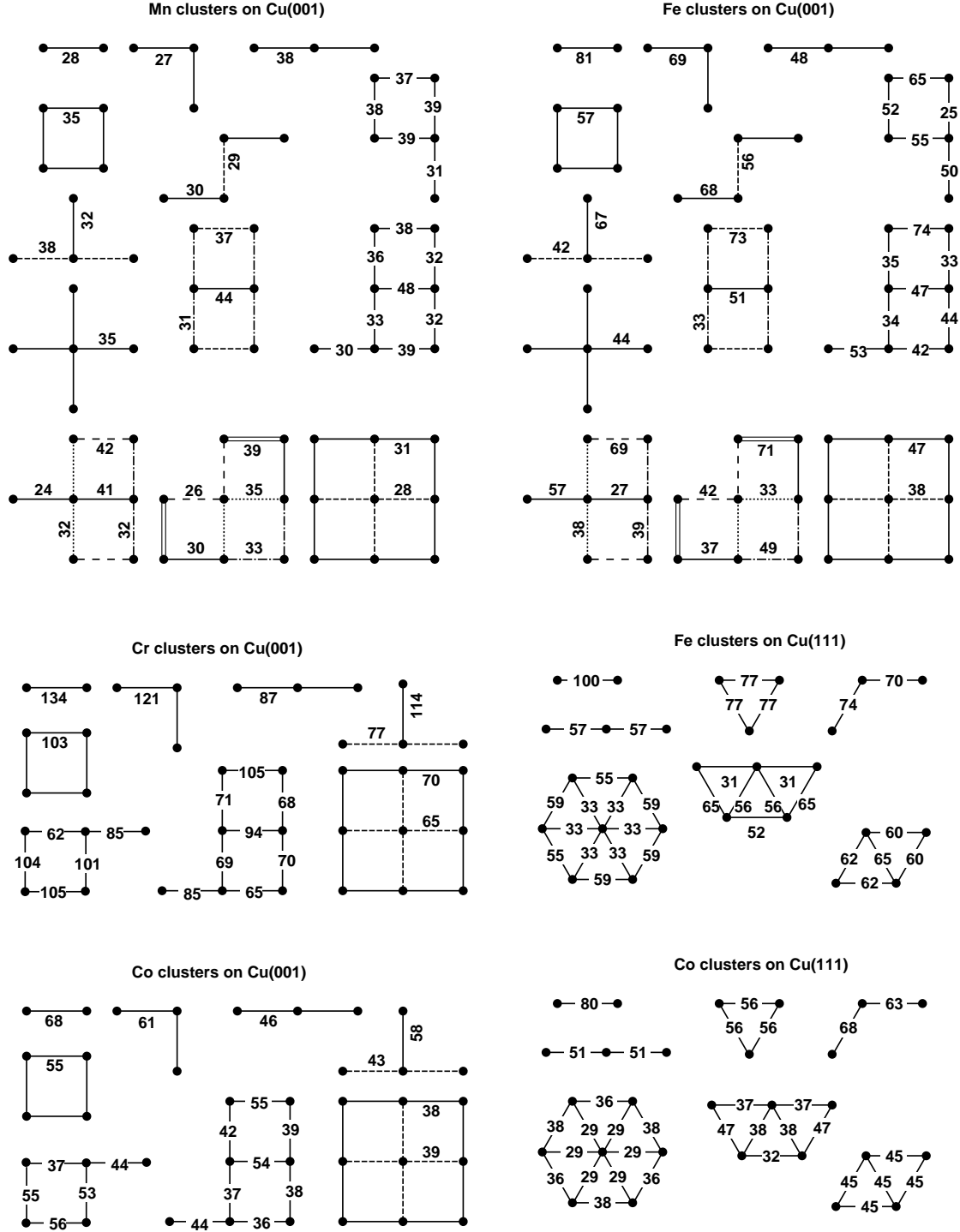}
\caption{Nearest-neighbour exchange coupling (absolute value, in meV)
  for Cr, Mn, Co, and Fe nano-clusters on Cu(001) and Cu(111). A
  schematic top-view of the clusters is shown, the Cu surface is
  understood to be one atomic layer below. The view corresponds to
  surface-adapted geometry, i.e., rotated with respect to the fcc
  cubic axes by $45^{\circ}$. Coupling values identical by symmetry
  are not annotated more than once, but indicated by the same
  line-type (e.g., full or dashed line). For Co/Cu(001), Cr/Cu(001),
  and Fe and Co/Cu(111), only selected clusters are shown. On the
  (111) surface, couplings that appear on first sight equivalent by
  symmetry can differ, due to the relative displacement of the
  sub-surface layer. \label{fig:jij}}
\end{center}
\end{figure*}

We now present and analyze the exchange coupling calculated within
density-functional theory. Results on most of the investigated
clusters are shown in Fig.~\ref{fig:jij}. As we expected, we find an
antiferromagnetic coupling for Cr and Mn clusters, dominated by
kinetic exchange, while Fe and Co are ferromagnetic, dominated by
double exchange. On the average, the interaction strength decreases
with increasing cluster size, dimers showing the strongest coupling,
but there are many exceptions (especially for Mn), while the local
geometry around each atom plays a role.

Cr is the champion in the strength of exchange coupling, with values
as high as $|J|=134$~meV for the Cr/Cu(001) dimer. Mn clusters show
comparatively weak coupling: for the Mn/Cu(001) dimer we find
$|J|=28$~meV, a value which increases by up to 70\% for certain
configurations. In Fe the coupling is again strong, reaching
$J=81$~meV for the Fe/Cu(001) and 100~meV for the Fe/Cu(111) dimer. In
Co, finally, the coupling is again relatively weaker, reaching values
of $J=68$~meV for the Co/Cu(001) and 80~meV for the Co/Cu(111)
dimer. We proceed with an analysis of our results based on the model
that were presented in the preceding section.

\subsection{Dimers: exchange coupling and density of states}  

In a first step, we discuss the exchange coupling in dimers on Cu(001)
and relate it to the electronic structure. The picture that is derived
here is helpful in understanding the behavior of larger clusters.

Our guide for the discussion is Fig.~\ref{fig:dos}. There, the
the density of states is shown for one of two atoms of a
Cr, Mn, Fe, and Co dimer (the other atom has exactly the same DOS, but
for Cr and Mn the spin directions are interchanged). In the same
figure, the energy-dependent exchange coupling density, $j(E)$, and
coupling, $J(E)=\int^{E} j(E') dE'$, are shown (see Eq.~\ref{eq:4}).

In all cases, the majority- and minority-spin $d$-resonances can be
clearly seen. Around each resonance, $j(E)$ (dash-dotted red line)
shows an S-shaped form, revealing the change of the resonance form
upon rotating the spins with respect to each other. In Cr and Mn, the
lower levels of each resonance will move even lower and the higher
levels will move higher: the resonance will widen up. This is because
the starting point is here an antiferromagnetic orientation of the
moments, and rotation allows for majority-majority and
minority-minority hopping. In Fe and Co, where the starting point is a
ferromagnetic orientation, the opposite will happen: the resonance
will narrow down, because majority-majority and minority-minority
begins to be blocked; thus, the S-shape is inverted. As the minority
spin resonance is bisected by \EF\ for Fe and Co, the single-particle
part of the energy, $\int^{\EF}E\,n(E)\,dE$, will increase, signalling
a ferromagnetic coupling of the double-exchange type. Note that the
behaviour of $j(E)$ is more spiked at the majority resonance in Mn and
Fe, due to the hybridization with the Cu $d$-states; moreover, in Cr,
Mn, and Fe the shape of $j(E)$ around the minority-spin resonance
appears to have a double-S form, revealing contributions from
different $d$-orbitals at slightly different energies. Around the
S-shapes, the integral $J(E)$ shows peaks (positive or negative,
depending on the starting moment configuration) revealing the total
strength of the interaction.

At some energy between the resonances, $j(E)$ passes through zero;
however, $J(E)$ can show a plateau with a local maximum at those
energies. This reveals a net shift of the resonances, apart from the
broadening or narrowing, upon rotation of the moments. The net shift
stems from the majority-minority-state repulsion, and can be
identified with the kinetic exchange. Its maximal value, indicated by
arrows in Fig.~\ref{fig:dos}, is strong in Cr, but less so in Mn,
probably because the Mn $d$-states are lower in energy and thus more
localized spatially, reducing the hopping. This is one reason why Mn
shows a weaker exchange coupling than Cr. Another, more important
reason is that for Cr \EF\ bisects $J(E)$ at a high point, in the
middle of the plateau, while for Mn \EF\ is already at a point strong
descent of $J(E)$, close to changing sign. Here, the double-exchange
mechanism among the minority-spin states is already starting to set
in, competing with the kinetic exchange.\cite{Mokrousov07} This
affects the behaviour of the coupling with increasing cluster size, as
we discuss later.

The kinetic exchange is also present for Fe and Co; it is just that
double-exchange is stronger and dominates, causing a ferromagnetic
ground state. It is interesting to disclose the strength of the
double-exchange by shifting the majority-spin resonances of Fe and Co
to lower energies, so that the kinetic exchange is diminished. We
achieved this by acting on these states with an attractive potential
of 2.7~eV. This resulted in an increase of the coupling by
approximately 50\% for all Fe and Co clusters. An energy-resolved
picture of this can be seen in the case of a Fe dimer, the DOS of
which is shown in Fig.~\ref{fig:dos}, labelled ``Fe-de''. Here, $J(E)$
does not show a peak between the resonances any more (the kinetic
exchange is now absent), while it is bisected by \EF\ much closer to
its maximum than in the normal Fe dimer.

\subsection{Trimers: the role of local geometry}

Having discussed the exchange in dimers, we now focus on the
complications brought about by local geometry effects. These can be
best demonstrated in the example of trimers, since for larger clusters
there are too many parameters that play a role.

We begin with an observation. The exchange coupling in linear trimers
is weaker than in corner-shaped ones (see Fig.~\ref{fig:jij}). This is
true for Cr/Cu(001), Fe and Co on Cu(001) and Cu(111), but not for
Mn/Cu(001). This behavior can be interpreted again in terms of a
tight-binding model, but with more than one orbitals. We will do so,
deferring the explanation of the Mn to a later point in this subsection.

\begin{figure*}
\begin{center}
  \includegraphics[width=0.9\textwidth]{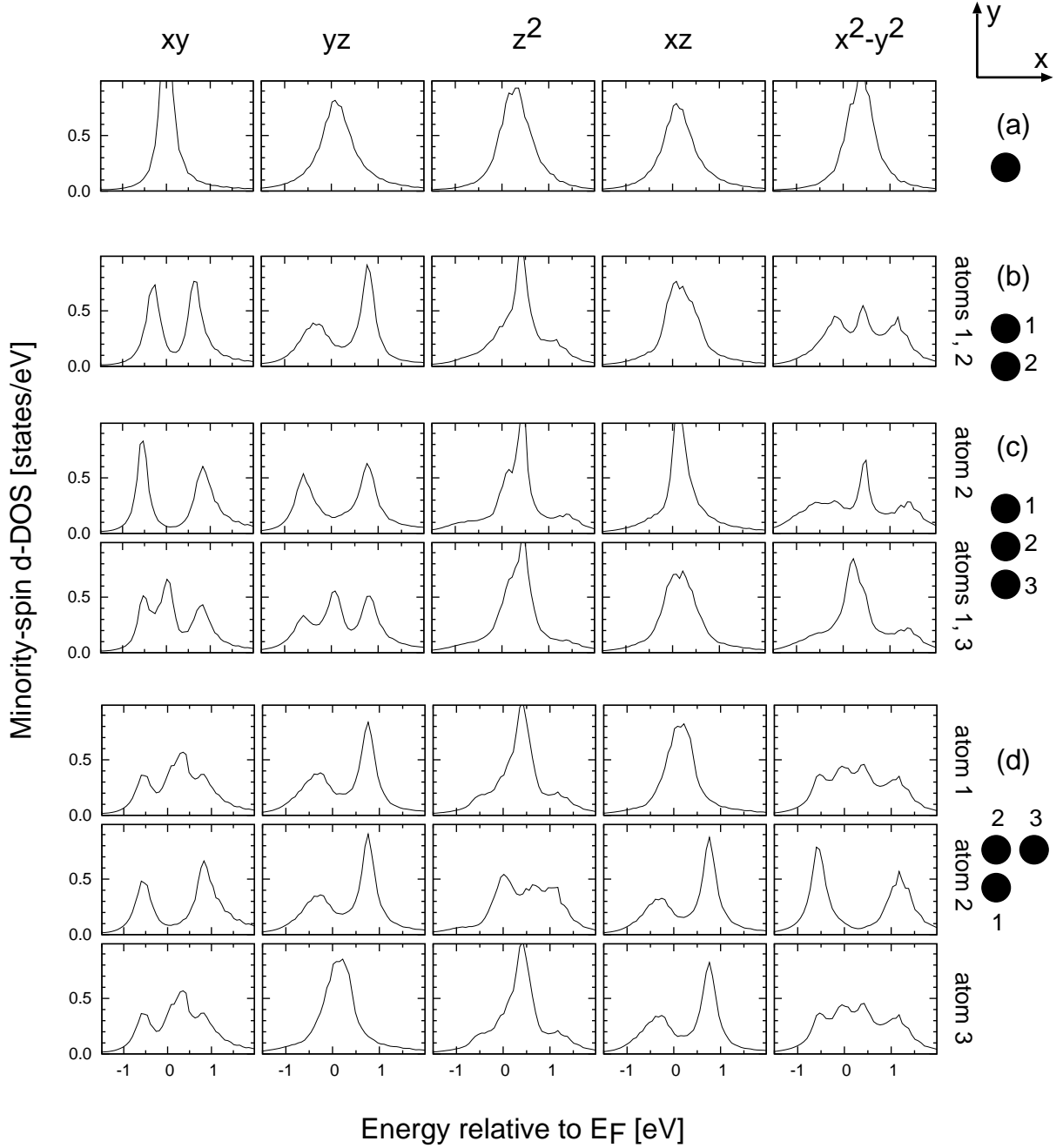}
  \caption{$d$-density of states resolved in $m$ for a Fe monomer (a),
    dimer (b), linear trimer(c) and corner-shaped trimer (d). The resonance
    splittings determine the energy gain from the ferromagnetic state formation.
    \label{fig:lmdos}}
\end{center}
\end{figure*}

In the interaction causing the kinetic or double exchange, not all $d$
orbitals take part equally. Some can have an orientation not favouring
bonding, while others can be pointing directly into the neighbouring
atom. Consider for instance a linear-shaped trimer in the
$y$-direction, setting for definiteness the $x$ and $y$ axes in the
surface plane (schematic positioning of atoms 1, 2, 3 in
Fig.~\ref{fig:lmdos}c). The \dxy, \dyz, and \dxxyy\ orbitals point
partly toward the neighbour atoms and can hybridize with their
counterparts there. The \dzz\ orbitals point out-of-plane, and can
therefore hybridize only weakly with \dxxyy\ neighbouring orbitals
(hybridization with the rest is forbidden by symmetry, if one
disregards the surface-induced symmetry breaking in a first
approximation). Finally, the \dxz\ orbitals point away from the
neighbours, and are practically non-bonding. Thus three channels have a
major contribution to the exchange coupling: one consists of the three
\dxy\ orbitals, one stems from the three \dyz, and one from the three
\dxxyy. Because each of these channels is brought about by three
orbitals, we call them ``triple channels'' henceforth. If the trimer
were to be extended to a linear chain, these three triple channels
would form the most dispersive $d$ bands.

Now let us change the picture to a corner-shaped trimer (schematic
positioning of atoms 1, 2, 3 in Fig.~\ref{fig:lmdos}d). Here we have
only two triple channels, namely the \dxxyy\ and the \dxy\
channels where all atoms participate. However, we also have two bonds:
\dyz-\dyz\ between atoms 1 and 2, and \dxz-\dxz\ between atoms 2 and
3. In effect, compared to the linear trimer, one triple channel has
been replaced by two bonds.

A straightforward calculation shows that the energy gain of two simple
bonds is larger than the one of a triple channel. Considering first
the case of double exchange, assume that the majority-spin $d$ levels
of the three atoms lie at $E_d+\Delta = \EF = 0$. In the linear
trimer, if we open a triple channel through the \dyz\ states by
allowing for a hopping $t$, a threefold splitting is created with
$E_{\pm}=\sqrt{2}t$, $E_0=0$; the \dxz\ orbitals are not affected. At
half-filling of the minority-spin orbitals (optimal case for
double-exchange), the energy gain is $\Delta E_{\rm
  tot}^{\text{Linear}} = \sqrt{2}t$. In the corner-shaped trimer,
allowing a hopping between \dyz-\dyz\ of atoms 1 and 2 and \dxz-\dxz\
of atoms 2 and 3 creates two double splittings of $E_{\pm}=t$. Now, at
half-filling of the minority-spin orbitals, the energy gain is $\Delta
E_{\rm tot}^{\text{Corner}} = 2t$. Evidently, $\Delta E_{\rm
  tot}^{\text{Corner}} >\Delta E_{\rm tot}^{\text{Linear}}$. Carrying
out the calculation for the kinetic-exchange case at full
majority-spin and empty minority-spin orbitals gives $\Delta E_{\rm
  tot}^{\text{Linear}} = 2(\sqrt{\Delta^2+t^2}-\Delta)$, $\Delta
E_{\rm tot}^{\text{Corner}} = 4(\sqrt{\Delta^2+t^2}-\Delta)$, i.e.,
again $\Delta E_{\rm tot}^{\text{Corner}} >\Delta E_{\rm
  tot}^{\text{Linear}}$. In both cases, the linear trimer should show
a smaller exchange $J$ than the corner-shaped one. Note that no
second-neighbour hopping is necessary in this interpretation. On the
(111) surface, the corner-shaped trimer is not in a $\phi=90^{\circ}$
configuration, but rather at $\phi=60^{\circ}$ or
$\phi=120^{\circ}$. Then the effect is still present, but to a lesser
extent as the hoppings satisfy partially a triple channel and
partially a bond (the hopping contributing to the triple channel is
$t\cos\phi$, the one contributing to the bond is $t\sin\phi$).

This simple picture can be recognized in the $m$-resolved density of
$d$-states. In Fig.~\ref{fig:lmdos} we show the minority-spin $d$-DOS
for a Fe/Cu(001) adatom, dimer, linear and corner-shaped trimer. In
panel (a), among the resonances of the single adatom, \dxy\ and
\dxxyy\ are clearly the most localized, as the lobes of these orbitals
extend mainly in-plane. Upon adding a neighbouring Fe atom (panel b),
clear-cut splittings are formed for all orbitals that point toward the
Fe neighbour. Even the \dzz\ orbital hybridizes somewhat with the
\dxxyy\ of the neighbour, causing a three-peak structure in
\dxxyy. Only the \dxz\ orbital remains oblivious to the presence of
the second atom, as it points in the wrong direction and is orthogonal
to any neighbouring orbital pointing in its way.

Next we add a third atom to form a linear trimer (upper and lower
panel c). Here, more splittings appear. In the two outmost atoms, 1
and 3, orbital \dxy\ shows now a threefold splitting, corresponding to
the three levels $E_{\pm}$ and $E_0$ that are formed within a triple
channel, as we discussed above; the same is true for \dyz. This is the
onset of band formation. However, the state corresponding to $E_0$ has
no weight in the middle atom, which can also be verified by solving
the associated $3\times 3$ system of equations within the
tight-binding model. For the \dxxyy\ orbitals the picture is more
complicated due to their additional hybridization with \dzz.

Finally, we move one atom to form a corner-shaped trimer (panel
d). Here the \dyz\ triple channel is broken. Instead, we have a bond
between the \dxy\ of atoms 1 and 2, and an identical bond between the
\dxz\ of atoms 2 and 3. The splittings, looking exactly like the \dyz\
splitting of the dimer of panel (b), verify this picture. We conclude
that the tight-binding model indeed interprets the calculated DOS and
coupling in Fe and Co.

We turn now to the interpretation of the behavior of Cr and Mn. For
Cr, which is a kinetic-exchange system, it is not the level splittings
but the level repulsion that causes the $90^{\circ}$, corner-shaped
configuration to show a stronger coupling than the linear
configuration, as outlined above. Mn, on the other hand, where the
competition of kinetic and double exchange is important, behaves
differently. Upon going from the linear trimer to the corner-shaped
one, \emph{both} the kinetic and the double exchange are
strengthened. Apparently, however, the enhancement of the
latter is stronger, so that, in the end, they sum up to a weaker
antiferromagnetic coupling. 

To conclude this subsection, a simple, nearest-neighbour tight-binding
model can be used to qualitatively interpret many of the features of
the DOS in dimers and trimers. This supports the interpretation of the
exchange coupling that we developed earlier, without a need to
introduce second-neighbour corrections; such corrections are of course
always present, but apparently not of central importance. The
complications of local geometry effects are revealed and explained in
the case of trimers; in particular it becomes clear that very
important for the behavior is not only the homogeneous broadening of
the resonances, but also the onset of band formation creating
geometry-dependent splittings.  If the moments are tilted away from
parallel orientation, these energy-gaining splittings will start
closing up, giving rise to the exchange coupling.  We now proceed to
the discussion of the exchange in larger clusters.

\subsection{Increasing size: effect of hybridization} 

Except for the case of Mn, increasing cluster size leads to weaker
coupling. This effect is coming from the stronger hybridization with
increasing coordination. For Fe and Co, hybridization contributes to
the lowering of exchange coupling in two ways. First, the double-exchange
interaction is weakened, as we described in section
\ref{sec:tb}. Second, the tendency for magnetism is weakened, as
reflected also in the reduction of the atomic moments of Fe.

For Cr clusters, which are a kinetic-exchange systems, mainly
the latter effect should play a role, since kinetic exchange should
depend on the splitting but not on the hybridization, as we argued in
section \ref{sec:tb}. However, for this argument it was hypothesized
that the resonance does not cross \EF, which is not completely true
here. As one can see from the density of states (Fig.~\ref{fig:dos}),
the resonances do cross \EF\ to an extent that increases with
hybridization (reflected also in the decrease of the Cr moments). Thus
the repopulation of the resonance tails also contributes to the
weakening of the kinetic exchange.

The effect of hybridization becomes clearer if one compares the values
of the exchange coupling on Cu(001) to the ones on Cu(111). In the
former case, each atom in the cluster has four Cu neighbours, in the
latter only three. We see that, on the average, the coupling on
Cu(001) is indeed weaker than on Cu(111) (for the same number of
transition-metal neighbours). Further evidence comes from examining Fe
and Co clusters on Ni(001) and (111) surfaces (not shown here in
detail). Here, additional background hybridization is induced on the
one hand from the Ni $d$ states at \EF\ and on the other hand from the
smaller lattice parameter. Although the moments, assisted by the
ferromagnetic substrate, are slightly larger than on Cu,
\cite{Mavropoulos06} the exchange coupling within the cluster is
significantly weaker, by a factor of up to 50\% for Fe/Ni(001)
compared to Fe/Cu(001), 40\% for Fe/Ni(111) compared to Fe/Cu(111),
and 25\% for Co/Ni(001) compared to Co/Cu(001).

It is also interesting to compare the systems Co/Pt(111) to
Co/Au(111), calculated by \v{S}ipr et al.\cite{Sipr07} In order to
eliminate the effect of increased lattice parameter, \v{S}ipr at
al. also calculated Co/Au(111) in the lattice parameter of
Pt. Although the Pt substrate is polarizable, enhancing the moments of
Co/Pt compared to Co/Au, the pair exchange coupling is weaker for
Co/Pt dimers and trimers, compared to the same formations on
Au.\cite{Siprprivate} This effect could be due to the strong
hybridization of the Co $d$ states with the Pt $d$ bands at \EF\ (the
$d$ bands of Au are considerably lower). For larger clusters, however,
this trend is not so clear and depends on the local
geometry. Additional evidence on the effect of hybridization comes
from calculations of Co ad-clusters on Pt(111) in
Ref.~\onlinecite{Minar06} where it is found that the nearest-neighbour
exchange of Co atoms at the cluster surface are larger by
approximately a factor 2 compared to interactions of a central Co
atom.

Finally we comment on the irregular behaviour of Mn clusters. The
irregularity arises from the competition of the kinetic and double
exchange analyzed in the discussion on the dimers above, and we also
saw it in the case of trimers. Small changes in the electronic
structure, caused e.g. by local geometry effects or cluster size, can
shift the turning point of $J(E)$ up and down in energy, disturbing
the delicate balance between kinetic and double exchange. This is the
reason that the Mn clusters show a rather irregular behavior in
$J_{nn'}$. More to this point, increasing coordination and
hybridization will suppress the double-exchange mechanism stronger
than the kinetic exchange, as we discussed in
Sec.~\ref{sec:tb}. Therefore, the exchange coupling tends to increase
with size in Mn clusters, opposite to Cr, Fe, and Co.

\subsection{Correlation of exchange and formation energy}

\begin{figure}[t]
\begin{center}
\includegraphics[width=8cm]{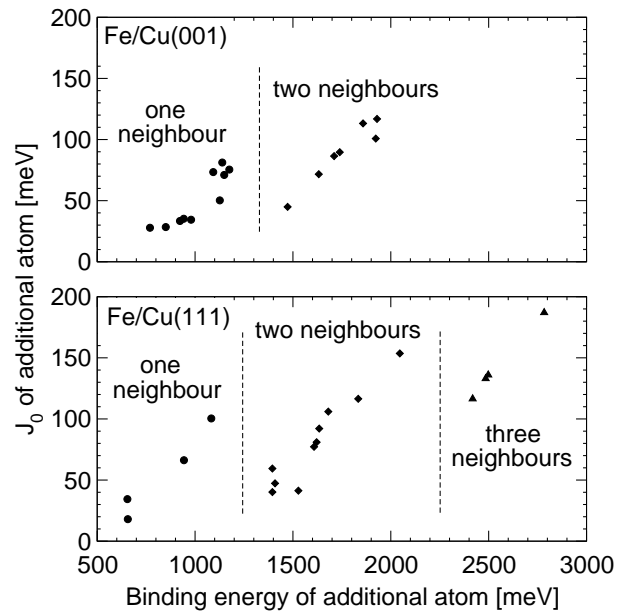}
\caption{Correlation between binding energy $|E_b|$ of a Fe atom to a
  Fe cluster on Cu(001) and Cu(111) vs. exchange coupling $J_0$ of the
  atom to the cluster. Dashed lines indicate the separation between
  atoms with one, two and three Fe neighbours.\label{fig:corr}}
\end{center}
\end{figure}

As we discussed so-far, level splittings and shifts are responsible
for the nearest-neighbour exchange coupling. However, it is also known
that such splittings are contributing to the binding
energy. Particularly for $d$-states, this idea dates back at least to
the model of Friedel for the cohesive energy of transition
metals. Therefore, there could be a correlation between the energy
gain upon attaching an atom to a cluster and the exchange coupling of
this atom with its neighbours. However, other factors also enter the
formation energy, and they could overweigh the effect of the $d$
states.

Of course there is a trivial correlation, in the sense that the total
exchange coupling energy of an atom increases with the number of
neighbours, and so does its binding energy. What we mean here is a
correlation even for atoms of the same coordination. To be more
precise, we are looking for a linear relation, of the form 
\begin{equation}
E_b(n) = A\sum_{n'}J_{nn'} + E_{\text{rest}}
\label{eq:jb}
\end{equation}
where $E_b(n)$ is the binding energy of the $n$-th atom to the rest of
the cluster and $A$ and $E_{\text{rest}}$ are constants.

In fact we do find such a correlation for Fe/Cu(001) and Fe/Cu(111),
but not for Cr, Mn, or Co (apart from the trivial one mentioned
above). We calculate the binding energy $E_b=E_{N+1} -
(E_{N}+E_{\text{ad}})$, where $E_N$ is the total energy of an $N$-atom
cluster, $E_{N+1}$ the total energy of the same cluster augmented by
one atom, and $E_{\text{ad}}$ the total energy of a lone-standing adatom
(total energies are calculated with respect to the reference energy of
the clean Cu surface).

In Fig.~\ref{fig:corr} we show the relation between the absolute value
of the binding energy of cluster atoms, $|E_b(n)|$, and the total
exchange coupling of these atoms to the rest of the cluster,
$J_0^{(n)}=\sum_{n'}J_{nn'}$. We only show results for atoms at the
cluster edge, as we consider it unlikely that a cluster will be formed
with an atom missing in the middle; therefore the maximal number of
neighbours is 3. We see a clustering of the data according to the
number of neighbours of the added atom, as expected, but within each
of these groups, a correlation of the form (\ref{eq:jb}) is
evident. There is also a clear ``background'' coupling, termed
$E_{\text{rest}}$ in relation (\ref{eq:jb}), which varies with the
number of neighbours. Deviations from relation (\ref{eq:jb}) can be
caused by several effects, such as participation of the $s$ electrons
in the bonding, charge relaxation, moment relaxation (which is rather
strong for Cr and Mn), etc. In addition, in the case of Mn, Fe and Co,
double exchange and kinetic exchange compete in the final value of
$J_{nn'}$, while the level shifts associated with kinetic exchange do
not contribute to the binding energy for Fe and Co, and the shifts
associated with double exchange do not contribute to binding in Mn.

\section{Conclusions}

The exchange coupling in supported transition metal nano-clusters
shows strong fluctuations, depending on cluster size, shape, and
position of the atoms in the cluster. Nevertheless, trends can be
observed, and the coupling can be analyzed and understood with respect
to simple physical terms, namely ferromagnetic double-exchange,
antiferromagnetic kinetic exchange, hybridization of the $d$-states,
and orientation of the $d$-orbitals. As these mechanisms act in
parallel, they interfere and compete, giving at the end a complex and
seemingly irregular picture. It becomes then difficult to isolate the
mechanisms in experiment for a direct testing of the theory.

We found that the physical picture and the trends become much more
transparent in simple cases, e.g., when studying dimers or trimers of
various configurations. We believe therefore that best insight can be
gained by studying highly-symmetric, low-dimensional structures, where
only one or two parameters define the structure. Such could be, for
instance, linear vs. corner-shaped trimers, or linear vs. zig-zag
chains with varying length. As there is considerable progress in
experimental techniques for preparation and probing of such
structures, a detailed and direct comparison with theoretical results
should be possible in the near future.

\acknowledgments 

We are indebted to Ondrej \v{S}ipr for sending us detailed results on
the Co/Au and Co/Pt systems, and to Peter Dederichs and Yuriy
Mokrousov for enlightening discussions. Financial support from the
Deutsche Forschungsgemeinschaft (DFG), within the Priority Programme
``Clusters in contact with surfaces: electronic structure and
magnetism'' (SPP-1153), is gratefully acknowledged.



\begin{thebibliography}{99}

\bibitem{Lau02}
J.T. Lau, A. F\"ohlisch, R. Nietuby\`{c}, M. Reif, and W. Wurth 
Phys. Rev. Lett. {\bf 89}, 057201 (2002).

\bibitem{Gambardella03}
P.~Gambardella, S.~Rusponi, M.~Veronese, S.S.~Dhesi, C.~Grazioli,
A.~Dallmeyer, I.~Cabria, R.~Zeller, P.H.~Dederichs, K.~Kern,
C.~Carbone, and H.~Brune, Science {\bf 300}, 1130 (2003).


\bibitem{Wildberger95}
K.~Wildberger, V.S.~Stepanyuk, P.~Lang, R.~Zeller, and P.H.~Dederichs,
Phys. Rev. Lett. {\bf 75}, 509 (1995); 
V.S.~Stepanyuk, W. Hergert, K.~Wildberger, S. K. Nayak, and P. Jena,
Surf. Sci. {\bf 384}, L892 (1997);
V. S. Stepanyuk, W. Hergert, P. Rennert, B. Nonas, R. Zeller, and
P. H. Dederichs, Phys. Rev. B {\bf 61}, 2356 (2000).

\bibitem{Stepanyuk99}
V. S. Stepanyuk, W. Hergert, P. Rennert, K. Wildberger, R. Zeller, and
P. H. Dederichs,
Phys. Rev. B {\bf 59}, 1681 (1999).


\bibitem{Izquierdo00}
J. Izquierdo, A. Vega, L. C. Balbas, D. Sanchez-Portal, 
J. Junquera, E. Artacho , J. M. Soler, and P. Ordejon,
Phys. Rev. B {\bf 61}, 13639 (2000); 

\bibitem{Spisak02}
D. Spi\v{s}\'ak and J. Hafner, Phys. Rev. B {\bf 65}, 235405 (2002);
Claude Ederer, Matej Komelj, and Manfred F\"ahnle,
Phys. Rev. B {\bf 68}, 052402 (2003);
A. B. Shick, F. M\'aca, and P. M. Oppeneer, Phys. Rev. B {\bf 69}, 212410 (2004).


\bibitem{Lazarovits02}
B. Lazarovits, L. Szunyogh, and P. Weinberger,
Phys. Rev. B {\bf 65}, 104441 (2002).

\bibitem{Mavropoulos06}
P. Mavropoulos, S. Lounis, R. Zeller, and S. Bl\"ugel, Appl. Phys. A
{\bf 82}, 103 (2006).


\bibitem{Hafner07}
J. Hafner and D. Spisak, Phys. Rev. B {\bf 76}, 094420 (2007).


\bibitem{Sipr07}
O. \v{S}ipr, S. Bornemann, J. Min\'{a}r, S. Polesya, V. Popescu,
A. Simunek and H. Ebert,
J. Phys.: Condens. Matter {\bf 19}, 096203 (2006).

\bibitem{Robles08}
R. Robles, A. Bergman, A. B. Klautau, O. Eriksson and L. Nordstr\"om 
J. Phys.: Condens. Matter {\bf 20}, 015001 (2008).


\bibitem{Etz07}
C. Etz, B. Lazarovits, J. Zabloudil, R. Hammerling, B. Ujfalussy,
L. Szunyogh, G. M. Stocks, and P. Weinberger, Phys. Rev. B {\bf 75},
245432 (2007).

\bibitem{Bergman06}
A. Bergman, L.Nordstr\"om, A. B. Klautau, S. Frota-Pessoa, and O. Eriksson
Surface Science {\bf 600}, 4838 (2006).


\bibitem{Bergman07}
A. Bergman, L. Nordstr\"om, A. B. Klautau, S. Frota-Pessoa, and
O. Eriksson  J. Phys.: Condens. Matter {\bf 19}, 156226 (2007).


\bibitem{Lounis05}
S. Lounis, Ph. Mavropoulos, P. H. Dederichs, and S. Bl\"ugel
Phys. Rev. B {\bf 72}, 224437 (2005).

\bibitem{Lounis07}
S. Lounis, Ph. Mavropoulos, R. Zeller, P. H. Dederichs,
and Stefan Bl\"ugel, Phys. Rev. B {\bf 75}, 174436 (2007).

\bibitem{Lounis08a}
S. Lounis, P.H. Dederichs, and S. Bl\"ugel,
Phys. Rev. Lett. {\bf 101}, 107204 (2008).

\bibitem{Lounis08b}
S. Lounis, M. Reif, P. Mavropoulos, L. Glaser, P. H. Dederichs,
M. Martins, S. Bl\"ugel and W. Wurth Europys. Lett. {\bf 81}, 47004
(2008).



\bibitem{Hirjibehedin07}
C. F. Hirjibehedin, C.-Y. Lin, A. F. Otte, M. Ternes, C. P. Lutz,
B. A. Jones, and A. J. Heinrich, Science {\bf 317}, 1199 (2007).



\bibitem{Otte08}
A. F. Otte, M. Ternes, K. von Bergmann, S. Loth, H. Brune, C. P. Lutz,
C. F. Hirjibehedin, and A. J. Heinrich, Nature Physics {\bf 4}, 847
(2008). 

\bibitem{Gao08}
C. L. Gao, A. Ernst, G. Fischer, W. Hergert, P. Bruno, W. Wulfhekel,
and J. Kirschner, Phys. Rev. Lett. {\bf 101}, 167201 (2008).

\bibitem{Balashov08}
T. Balashov, A. F. Takacs, M. Dane, A. Ernst, P. Bruno, and
W. Wulfhekel, Phys. Rev. B {\bf 78}, 174404 (2008).


\bibitem{Meier08}
F. Meier, L. Zhou, J. Wiebe, and R. Wiesendanger,
Science {\bf 320}, 82 (2008).


\bibitem{Minar06}
J. Min\'{a}r, S. Bornemann, O. \v{S}ipr, S. Polesya, and H. Ebert,
Appl. Phys. A {\bf 82}, 139 (2006).



\bibitem{Polesya07}
S. Polesya, O. \v{S}ipr, S. Bornemann, J. Min\'{a}r, and H. Ebert,
Europhys. Lett. {\bf 74} 1074 (2006).


\bibitem{Vosko80}
S.H. Vosko, L. Wilk, and M. Nusair, Can. J. Phys. {\bf 58}, 1200 (1980).

\bibitem{Papanikolaou02}
N. Papanikolaou, R. Zeller and P.H. Dederichs
J. Phys.: Condens. Matter {\bf 14}, 2799 (2002);
N. Stefanou, H. Akai, and R. Zeller, Comp. Phys. Commun. {\bf 60}, 231 (1990);
N. Stefanou and R. Zeller, J. Phys.: Condens. Matter {\bf 3}, 7599
(1991).


\bibitem{Liechtenstein87}
A.I. Liechtenstein, M.I. Katsnelson, V.P. Antropov, and V.A. Gubanov,
J.~Magn.~Magn.~Mater.~{\bf 67} 65 (1987).

\bibitem{Pajda02}
M. Pajda, J. Kudrnovsk\'y, I. Turek, V. Drchal, and P. Bruno,
Phys. Rev. B {\bf 64}, 174402 (2001).


\bibitem{Brovko09}
O. O. Brovko, W. Hergert, and V. S. Stepanyuk
Phys. Rev. B 79, 205426 (2009).

\bibitem{Brovko08}
O. O. Brovko, P. A. Ignatiev, V. S. Stepanyuk, and P. Bruno,
Phys. Rev. Lett. {\bf 101}, 036809 (2008).



\bibitem{Weismann09}
A. Weismann, M. Wenderoth, S. Lounis, P. Zahn, N. Quaas,
R. G. Ulbrich, P. H. Dederichs, and S. Bl\"ugel Science {\bf 323},
1190 (2009). 


\bibitem{Bode07}
M. Bode, M. Heide, K. von Bergmann, P. Ferriani, S. Heinze, G. Bihlmayer, A. Kubetzka, O. Pietzsch, S. Bl\"ugel, and R. Wiesendanger, Nature (London) 447, 190 (2007)

\bibitem{Antal08}
A. Antal, B. Lazarovits, L. Udvardi, L. Szunyogh, B. Ujfalussy, and
P. Weinberger, Phys. Rev. B {\bf 77}, 174429 (2008).

\bibitem{Mankovsky09}
S. Mankovsky, S. Bornemann, J. Minar, S. Polesya, H. Ebert,
J. B. Staunton, and A. I. Lichtenstein, Phys. Rev. B {\bf 80}, 014422 (2009).


\bibitem{Alexander64}
S. Alexander and P. W. Anderson 
Phys. Rev. {\bf 133}, A1594 (1964).

\bibitem{Akai98}
The same mechanisms appear and compete also in diluted magnetic semiconductors;
see, e.g., 
H. Akai, Phys. Rev. Lett. {\bf 81}, 3002 (1998);
B. Belhadji, L. Bergqvist, R. Zeller, P.H. Dederichs, K. Sato, and
H. Katayama-Yoshida, J. Phys.: Condens. Matter {\bf 19}, 436227
(2007).


\bibitem{Mokrousov07}
Also monoatomic Cr chains show stronger antiferromagnetism than Mn
chains. See, e.g., 
Y. Mokrousov, G. Bihlmayer, S. Bl\"ugel, and S. Heinze,
Phys. Rev. B 75, 104413 (2007).


\bibitem{Siprprivate}
O. \v{S}ipr, private communication.



\end{thebibliography}
\end{document}